\documentclass[aps, pra, amsfonts, amssymb, amsmath, showkeys, twocolumn, nofootinbib, 10pt]{revtex4-2}
\usepackage{lmodern}

\usepackage[english]{babel}
\usepackage[T1]{fontenc}
\usepackage[utf8]{inputenc}
\usepackage[bottom]{footmisc}
\usepackage{physics}
\usepackage{esint}
\usepackage{color}
\usepackage{graphicx}
\usepackage{geometry}
\usepackage{float}
\usepackage{booktabs}
\usepackage[unicode=true,pdfusetitle,
bookmarks=true,bookmarksnumbered=false,bookmarksopen=false,
 breaklinks=false,pdfborder={0 0 1},backref=false,colorlinks=true]
 {hyperref}
\hypersetup{citecolor=blue,linkcolor=blue,urlcolor=blue}
\usepackage{float}
\usepackage[utf8]{inputenc}

\begin{document}

\title{Constraints on Quintom Dark Energy Model with Matter Interaction}
\author{Chakkrit Kaeonikhom}
    \email{chakkrit_kae@g.cmru.ac.th}
    \affiliation{Department of Physics and General Science, \\ Faculty of Science and Technology, Chiang Mai Rajabhat University, Mae Rim Campus, Mae Rim Dist., Chiang Mai 50180, Thailand}
\author{Theppawan Rukpakawong}
    \email{theppawan.k@gmail.com}
    \affiliation{High Energy Physics Theory Group, Department of Physics, Faculty of Science, Chulalongkorn University, Bangkok 10330, Thailand}
\author{Piyabut Burikham}
    \email{piyabut@gmail.com}
    \affiliation{High Energy Physics Theory Group, Department of Physics, Faculty of Science, Chulalongkorn University, Bangkok 10330, Thailand}
\author{}
\date{\today}

\begin{abstract}

The quintom model consisting of quintessence and phantom scalar fields with linear coupling to matter is explored and constrained by updated DESI and supernovae type Ia~(SNIa) observational data. The matter interaction is assumed only between phantom and matter and the quintessence self potential is in an exponential form. The quintom evolution shows a class of solutions with phantom crossing scenario consistent with the current DESI data. From low redshift observations, we place constraints on the model with BAO data from DESI and distance measurements from SNIa Pantheon+ samples and SH0Es. For high-redshift constraints, we use CMB priors from Planck and found that the quintom model gives the best-fit Hubble parameter $H_{0}=65.780~(68.736)$~km s$^{-1}$ Mpc$^{-1}$ for DESI+CMB+SNIa~(+SH0ES) respectively. The marginalized statistical constrained values are $H_{0}=66.21\pm 0.84~(68.49^{+0.46}_{-0.40})$~km s$^{-1}$ Mpc$^{-1}$ for DESI+CMB+SNIa~(+SH0ES) respectively. Using the BBN prior for high-redshift constraints, the corresponding best-fit values for DESI+BBN+SNIa~(+SH0ES) constraints are $H_{0}=63.999~(71.962)$~km s$^{-1}$ Mpc$^{-1}$ and the marginalized statistical values for DESI+BBN+SNIa~(+SH0ES) constraints are $H_{0}=61.6^{+6.6}_{-3.0}~(71.86\pm 0.88)$~km s$^{-1}$ Mpc$^{-1}$ respectively. Generally, adding SH0ES significantly raises the best-fit and marginalized value of $H_{0}$ for all datasets. In comparison to $\Lambda$CDM, the interacting quintom model gives lower values of $\sigma_{8}=0.74-0.79$ for combination of datasets with CMB prior. Lastly, the fifth-force constraint from local scale solar system Cassini probe is estimated, we find the constraint on the phantom-matter coupling $|\delta|<3.3\times 10^{-3}$.  

 \end{abstract}

\maketitle

\section{Introduction} \label{sec:introduction}

Dark energy~(DE) and dark matter are among the most active fields of investigation and the most important problems in cosmology. Low redshift observational data from standard candles, i.e., Cepheid variable stars and Type Ia supernovae suggest accelerated expansion of the universe with Hubble parameter $H_{0}=73.30\pm 1.04$~km s$^{-1}$ Mpc$^{-1}$~\cite{riess2022comprehensive,scolnic2022pantheon+}. High redshift observation from the {\it Planck} Collaboration constrains the Hubble parameter to lower value $H_{0}=67.4\pm 0.5$~km s$^{-1}$ Mpc$^{-1}$~\cite{aghanim2020planck}. However, analysis of SNe Ia with a calibration of the tip of the red-giant branch gives a closer value to the high redshift result, $H_{0}=69.8\pm 2.2$~km s$^{-1}$ Mpc$^{-1}$~\cite{Freedman:2021ahq}. The method of time delay in gravitationally lensed quasar from a sample of 40 lenses also gives a relatively low value, $H_{0}=67.4^{+4.1}_{-3.2}$~km s$^{-1}$ Mpc$^{-1}$~\cite{Birrer:2020tax}. On the other hand, by using inverse distance ladder relies on absolute-distance measurements from baryon acoustic oscillations~(BAOs) to calibrate the intrinsic magnitude of the SNIa (instead of Cepheids), a combination of DES Year 3 (Y3) clustering and weak lensing with BAO and BBN (assuming $\Lambda$CDM) gives $H_{0}=67.6\pm 0.9$~km s$^{-1}$ Mpc$^{-1}$~\cite{DES:2021wwk}, a value close to the {\it Planck} result. 

Recent observational data from DESI DR2~\cite{tr6y-kpc6} statistically prefers phenomenological dynamical DE over the $\Lambda$CDM model. The data provides updated observational constraints from BAO which are collected at low redshifts and they should be consistent with low redshift constraints from SNIa observation. Best-fit $w_{0}w_{a}$-DE model of the DESI results has phantom-crossing characteristics, implying that the dynamical DE possibly has undergone a ``phantom phase'' during cosmological evolution. The dynamical DE then crossed the phantom divide $w_{DE}=-1$, and became non-phantom in the present era. 

In Ref.~\cite{Panpanich:2019fxq}, a quintom model with phantom-matter coupling is proposed which demonstrates phantom crossing behavior. It was found that for certain region of the parameter space, the coupled quintom model is more statistically preferred than the standard $\Lambda$CDM model. In this work, we revisit the coupled quintom model and extend our analyses to DESI results with updated analytic and numerical approach. We implement the model into CLASS to find the numerical cosmological solutions and use Cobaya for the MCMC~(Markov chain Monte Carlo) simulations under constraints from the observational data from DESI BAO, SNIa Pantheon+ and SH0ES, with BBN and CMB distance priors. In the presence of phantom-matter coupling, quintom model prefers cosmological evolution with $\Omega_{m}\simeq 0.30$ and $H_{0}=64.0-67.0$~km s$^{-1}$ Mpc$^{-1}$ for all combinations of datasets excluding SH0ES.

This work is organized as the following. Section \ref{sec:quintom_model} and \ref{sec:approx} establish the coupled quintom model with certain analytic results and discussion on approximation scheme. In Section \ref{sec:cmban}, we discuss effects of phantom-matter coupling in CMB anisotropy. Observational constraints are presented in comparison with the $\Lambda$CDM model in Section \ref{sec:obscon}. Cosmological inference and constraints to the models are discussed in Section \ref{sec:cosinf} and \ref{sec:constr}. Local scale constraints are explored in Section~\ref{SectCassini} and Section \ref{sec:con} concludes our work.  

\section{Quintom Dark Energy Model}\label{sec:quintom_model}
The quintom model is characterized by the presence of two distinct scalar fields with opposite kinetic signs. This unique configuration allows the model to traverse the \emph{phantom divide} $(w = -1)$, a feat theoretically impossible for a single canonical scalar field. The action for the quintom model is given by:
\begin{equation}
    S = \int \mathrm{d}^4x\sqrt{-g}\left[\frac{1}{2\kappa^2}R+\mathcal{L}_M+\mathcal{L}_Q\right],
\end{equation}
where $\kappa^2 = 8\pi G$ is an inverse of the reduced Planck mass squared, and $\mathcal{L}_Q$ represents the quintom Lagrangian:
\begin{equation}
    \mathcal{L}_Q = -\frac{1}{2}\partial_\mu\phi\partial^\mu\phi + \frac{1}{2}\partial_\mu\sigma\partial^\mu\sigma - V(\phi).
\end{equation}
Here, $\phi$ is the quintessence field (canonical) and $\sigma$ is the phantom field (non-canonical). The opposite sign in the kinetic term of $\sigma$ provides the necessary degrees of freedom for the total equation of state to enter the phantom regime $(w < - 1)$. 

We assume that there is only one self-interacting potential of the quintessence scalar field, while the phantom scalar field is rolling on an effective potential arising from the phantom-matter interaction. Moreover, this phantom scalar field is not exactly the standard phantom but rather a ghost field since its equation of state is $P_\sigma = \rho_\sigma < 0$. However, here and henceforth we will simply call it the phantom field for convenience, and also in accordance with the original name quintom (quintom + phantom).

By varying the action with respect to $g^{\mu\nu}$, using the flat
Friedmann–Lemaître–Robertson–Walker (FLRW) metric, and assuming $\phi = \phi(t)$ and $\sigma = \sigma(t)$, we obtain the Friedmann equations:
\begin{align}
    3H^2 &= \kappa^2(\rho_m+\rho_r+\rho_\phi+\rho_\sigma), \\
    3H^2 + 2\dot{H} &= \kappa^2(-P_m-P_r-P_\phi-P_\sigma),
\end{align}
where $\rho_m, \rho_r, P_m$ and $P_r$ are energy densities and pressures of non-relativistic matter and radiation, respectively. The notation " $\dot{}$ " means a derivative with respect to time. The energy densities and pressures of the scalar fields are 
\begin{align}
    \rho_\phi &= \frac{1}{2}\dot{\phi}^2 + V(\phi), \quad \rho_\sigma = -\frac{1}{2}\dot{\sigma}^2, \\
    P_\phi &= \frac{1}{2}\dot{\phi}^2 -V(\phi), \quad P_\sigma = -\frac{1}{2}\dot{\sigma}^2.
\end{align}

As we mention about the phantom-matter interaction, we consider the interaction in the form as follows~(proposed in Ref.~\cite{Panpanich:2019fxq}):
\begin{align}
    \nabla_\mu T^{\mu(M)}_\nu  &= \kappa\delta T_M \nabla_\nu\sigma, \\
    \nabla_\mu T^{\mu(\sigma)}_\nu &= - \kappa\delta T_M \nabla_\nu\sigma, \label{eq:phm}
\end{align}
where $T_M = -\rho_M + 3P_M$, and $\delta$ is a dimensionless constant. This is a conformal interaction form which arises in many scalar-tensor theories after taking a conformal transformation to the Einstein frame. Hence, the continuity equations are
\begin{align}
    \dot{\rho}_m + 3H\rho_m &= \kappa\delta\rho_m\dot{\sigma}, \label{eq:rhom} \\
    \dot{\rho}_{\sigma} + 3H(\rho_\sigma + P_\sigma) &= -\kappa\delta\rho_m\dot{\sigma}, \label{eq:rhosigma} \\
    \dot{\rho}_{\phi} + 3H(\rho_\phi+P_\phi) &= 0, \label{eq:rhophi}\\
    \dot{\rho}_{r} + 4H\rho_r &= 0. \label{eq:rhor}
\end{align}
Substituting energy density and pressure of each scalar field we find the equation of motion:
\begin{align}
    \ddot{\sigma}+3H\dot{\sigma} - \kappa\delta\rho_m &= 0 \label{eq:phantom_eq} \\
    \ddot{\phi}+3H\dot{\phi} + V,_{\phi} &= 0. \label{eq:scf_eq}
\end{align}
As we can see, $\kappa\delta\rho_m$ term acts as an effective potential. This is similar to the effective potential in the chameleon or symmetron gravity.

\section{Approximate analytic solutions}\label{sec:approx}

The continuity equations \eqref{eq:rhom}--\eqref{eq:rhor} can be rewritten as
\begin{align}
    \rho_m' + 3\rho_m &= \kappa \delta\, \rho_m\sigma',  \label{eq:rhom_conf} \\
    \rho_\text{de}' + 3(\rho_\text{de} + P_\text{de}) &= -\kappa \delta\, \rho_m \sigma', \\
    \rho_r' + 4\rho_r &= 0,
\end{align}
where the dark energy density, $\rho_\text{de}=\rho_\phi+\rho_\sigma$, is the combination of the quintessence and phantom fields, and the prime symbol denotes the differentiation with respect to the e-fold variable, $N\equiv\ln{a}$.
For a flat Universe, the Klein-Gordon equations for the phantom and quintessence fields, equations \eqref{eq:phantom_eq} and \eqref{eq:scf_eq} in e-fold number variable become
\begin{align}
    \sigma'' + \frac{3}{2}\left(1-w_\text{eff}\right)\sigma' - \kappa\delta\,\frac{\rho_m}{H^2} &= 0, \label{eq:KG_phantom} \\
    \phi'' + \frac{3}{2}\left(1-w_\text{eff}\right)\phi' + \frac{V,_{\phi}}{H^2} &= 0, \label{eq:KG_scf}
\end{align}
where $x'\equiv dx/dN$ and the effective equation of state, $w_\text{eff}\equiv P_{\rm total}/\rho_{\rm total}$ (for a flat Universe) can be expressed in the form
\begin{equation}
    w_\text{eff} = -1 - \frac{2}{3}\frac{H'}{H}.
\end{equation}
Equation \eqref{eq:rhom_conf} yields an analytic solution of the form
\begin{equation}
    \rho_m(a) = \rho_{m,0} a^{-3}e^{\kappa\delta\,(\sigma(a)-\sigma_0)},
\end{equation}
where the present-day matter density, given by
\begin{equation}
    \rho_{m,0} = K_m e^{\kappa\delta\,\sigma_0},
\end{equation}
is scaled by the phantom field's present value, $\sigma_0\equiv\sigma(a=1)$, and the coupling constant, $\delta$.
Since $\sigma$ has no potential, a constant shift of $\sigma$ can be absorbed into $K_m$. With the initial condition $\sigma_\text{ini}=0$ adopted below, $K_m=\rho_{m,0}\,e^{-\kappa\delta\sigma_0}$ is therefore fixed by the present-day matter density.
We note that in case of non-coupling, $\delta=0$, the constant $K_m$ becomes the present time matter density, i.e, $K_m=\rho_{m,0}$.
Although we obtain an exact solution for the matter density in this coupled quintom model, the phantom field $\sigma(a)$ must be evaluated numerically. Furthermore, since we assume the dark energy density is the combination of two fields, $\rho_\text{de}=\rho_{\phi}+\rho_{\sigma}$, solving for both fields requires specifying the quintessence field potential, $V(\phi)$.

In this model, we adopt the widely studied exponential potential for the quintessence field, $V(\phi)=V_0\,e^{-\kappa\lambda_\phi\phi}$, where $V_0$ is the potential amplitude and $\lambda_\phi$ is a dimensionless parameter characterizing the steepness of the potential.
%
We solve the background equations from an initial scale factor $a_\text{ini}=10^{-14}$, deep in the radiation era. The phantom field starts at rest, $\sigma_\text{ini}=0$ and $\dot\sigma_\text{ini}=0$. The choice $\sigma_\text{ini}=0$ involves no loss of generality: since $\sigma$ has no potential, a constant shift of $\sigma$ is absorbed into the constants $K_m$ of the coupled matter species. The quintessence field starts at $\phi_\text{ini}=1/\lambda_\phi$ with an initial kinetic energy $\tfrac12\dot\phi_\text{ini}^2=2V(\phi_\text{ini})$. Because $V(\phi_\text{ini})$ is of the order of the present dark-energy density, this kinetic energy is negligible compared with that of radiation, and it redshifts as $a^{-6}$; the field is therefore frozen long before nucleosynthesis, and its subsequent evolution is that of a field initially at rest. The value of $\phi_\text{ini}$ is likewise immaterial, since a constant shift of $\phi$ only rescales $V_0$. The amplitude $V_0$ is fixed by requiring the present-day dark-energy density to close the energy budget, $\rho_{\text{de},0}=3H_0^2/\kappa^2-\rho_{m,0}-\rho_{r,0}$, and the constants $K_m$ by the present-day densities of the coupled species. With these initial conditions the model reduces to $\Lambda$CDM, with $\Omega_\Lambda=\kappa^2V_0/(3H_0^2)$, when $\lambda_\phi=\delta=0$. 

To understand how this coupled quintom model deviates from the standard $\Lambda$CDM model, 
we analyze the scale factor at radiation-matter equality. Consider the ratio of radiation to matter density, given by
\begin{equation}
\frac{\rho_r(a)}{\rho_m(a)} = 
\frac{\rho_{r,0}a^{-4}}{\rho_{m,0} a^{-3}e^{\kappa\delta\,(\sigma(a)-\sigma_0)}}.
\end{equation}
Therefore the $a_\text{eq}$ is dynamically shifted by the phantom field at that epoch
\begin{equation}
    a_\text{eq} = \frac{\Omega_{r,0}}{\Omega_{m,0}}e^{-\kappa\delta\,(\sigma(a_\text{eq})-\sigma_0)},
\end{equation}
where $\Omega_{i,0}\equiv\kappa^2\rho_{i,0}/(3H_0^2)$ is the present-day density parameter. Deep in the radiation era, $a\ll a_\text{eq}$, the effective equation of state is $w_\text{eff}\simeq1/3$. The matter energy density is negligible compared to the total density, so $\rho_m / H^2 \approx 0$. The equation \eqref{eq:KG_phantom} simplifies to 
\begin{equation}
   \sigma''(N) + \sigma'(N) = 0. \label{eq:frozen_sigma}
\end{equation}
In equation \eqref{eq:frozen_sigma}, the source term vanishes and Hubble friction freezes the field at $\sigma(N) \approx \sigma_{\text{ini}} = \text{constant}$. 

For matter dominance, radiation is negligible, so $w_{\text{eff}} = 0$, and $\rho_m \approx \rho_{\text{tot}}=3H^2/\kappa^2$. The equation \eqref{eq:KG_phantom} becomes
\begin{equation}
    \label{eq:matter_d_om_sigma}
    \sigma''(N) + \frac{3}{2}\sigma'(N) = 3\delta/\kappa.
\end{equation}
During the matter-dominated era, the scalar field dynamics approach an asymptotic attractor solution. As Hubble friction balances the coupling source term, the field acceleration $\sigma''(N)$ in the equation \eqref{eq:matter_d_om_sigma} becomes negligible. In this limit, the equation of motion simplifies to $\sigma'(N) \approx 2\delta/\kappa$.
Consequently, the field velocity becomes constant with respect to e-foldings, resulting in a scalar field that evolves logarithmically with the scale factor, $\sigma \propto \ln a$. 
On this attractor, $\rho_\sigma=-\tfrac12H^2\sigma'^2$ and hence $\Omega_\sigma\to-2\delta^2/3$. The phantom energy density scales as matter, while the quintessence energy density stays frozen close to $V(\phi_\text{ini})$.

To approximate the phantom field value at radiation-matter equality, $\sigma_\text{eq} \equiv \sigma(a_\text{eq})$, we express the matter density in terms of $x \equiv a/a_\text{eq}$ as $\rho_m(x) = \rho_{m,\text{eq}} x^{-3} e^{\kappa \delta (\sigma(x) - \sigma_\text{eq})}$.
Assuming the field displacement $\Delta \sigma(x) \equiv \sigma(x) - \sigma_\text{eq}$ remains small near $\sigma(a_\text{eq})$, the ratio of matter to radiation energy densities simplifies to
\begin{equation}
    \frac{\rho_m(x)}{\rho_r(x)} = \frac{x^{-3} e^{\kappa \delta \Delta \sigma(x)}}{x^{-4}} = x e^{\kappa \delta \Delta \sigma(x)} \approx x.
\end{equation}
Here the prefactors cancel because $\rho_{m,\text{eq}}=\rho_{r,\text{eq}}$, and the last step requires $\kappa\delta\,\Delta\sigma\ll1$. 
Since dark energy is negligible at these redshifts, $w_\text{eff}=P_\text{tot}/\rho_\text{tot}\simeq(\rho_r/3)/(\rho_r+\rho_m)$. Consequently, the effective equation of state is approximately
\begin{equation}
    w_{\text{eff}}(x) \approx \frac{1}{3(1+x)},
\end{equation}
and the source term of \eqref{eq:KG_phantom} becomes $\rho_m/H^2 \approx 3x/[\kappa^2(1+x)]$. Substituting these approximations into equation \eqref{eq:KG_phantom} yields the approximate equation of motion
\begin{equation}
    \dv[2]{\sigma}{x} + \frac{5x+4}{2x(1+x)}\dv{\sigma}{x} = \frac{3\delta}{\kappa x(1+x)}.
\end{equation}
Integrating this equation with the boundary condition, $\dd{\sigma}/\dd{N}\to 0$ as $x\to 0$, which follows from $\dot{\sigma}_\text{ini}=0$ gives the field velocity
\begin{equation}
    \dv{\sigma}{x} = \frac{2\delta(1+x) - 6\delta + 4\delta/\sqrt{1+x}}{\kappa x^2}.
\end{equation}
%
Integrating from the deep radiation era, where $\sigma\to\sigma_\text{ini}$ as $x\to0$, to arbitrary $x$ gives $\kappa[\sigma(x)-\sigma_\text{ini}]/\delta=F(x)$, with
\begin{equation}
  F(x)\equiv4\ln\left[\frac{1+\sqrt{1+x}}{2}\right] + \frac{2\left(\sqrt{1+x}-1\right)}{\sqrt{1+x}+1}.
  \label{eq:helperFx}
\end{equation}
At equality, $x=1$, the total field displacement $\Delta\sigma_\text{eq}\equiv\sigma_\text{eq}-\sigma_\text{ini}$ is
\begin{equation}
    \Delta \sigma_\text{eq} = \frac{\delta}{\kappa} F(1) \approx 1.1\,\frac{\delta}{\kappa}\,.
\end{equation}
%
Since $F(x)$ does not depend on $\delta$, and $\dv*{F}{x}>0$, $F(x)$ increases monotonically from $F(0)=0$ to $F(1)\simeq1.1$. With $\Delta\sigma(x)=(\delta/\kappa)[F(x)-F(1)]$, this gives, for $0\le x\le1$,
\begin{equation}
  -1.1\,\delta^2\le\kappa\delta\,\Delta\sigma(x) \le0,
\end{equation}
whatever the sign of $\delta$. The approximation treats matter as conserved between $x$ and equality. Since the coupling feeds energy into matter, the exact ratio $\rho_m/\rho_r=x\,e^{\kappa\delta\,\Delta\sigma}$ lies below $x$, by a fraction that vanishes at equality and grows to at most $1.1\,\delta^2$ as $x\to0$.
Well after equality, where $x\gg1$, $F(x)$ in \eqref{eq:helperFx} can be approximated as 
\begin{equation}
  F(x)\simeq2\ln x+2-4\ln2+\frac{4}{x}+\mathcal{O}\!\left(x^{-3/2}\right).
\end{equation}
Hence $\sigma'=(\delta/\kappa)\,x\dv*{F}{x}\to2\delta/\kappa$, and the field joins the matter-era attractor. 
Once dark energy grows, both terms of the balance behind Eq.~\eqref{eq:matter_d_om_sigma} change. 
%

\begin{center}
\begin{figure}
    \includegraphics[width=\linewidth]{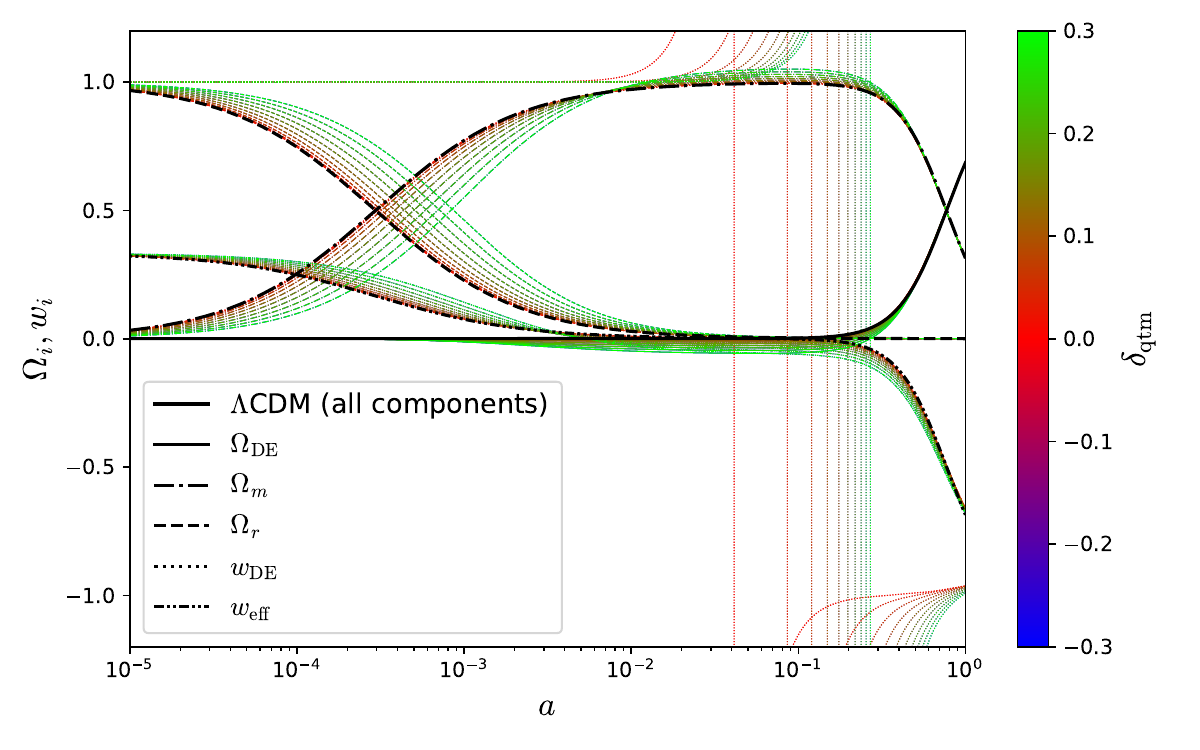}
    \caption{Evolution of the density parameters $\Omega_i$ and the equations of state $w_\text{DE}$ and $w_\text{eff}$ for several values of $\delta$ with $\lambda_\phi=0.5$, compared with $\Lambda$CDM. Curves for $\pm\delta$ coincide. Vertical lines mark the redshifts where $\rho_\text{de}=0$ and $w_\text{DE}$ diverges.}
    \label{fig:evol_densities}
\end{figure}
\end{center}

These results show that $\sigma'$, and hence the field displacement, is proportional to $\delta$.
Since $\sigma'$ starts at zero and is sourced by $\delta\rho_m$, it keeps the sign of $\delta$ at all times. The energy transfer $Q=\kappa\delta\rho_m\sigma'$ therefore scales as $\delta^2$ and
is non-negative for either sign of $\delta$, so energy flows only from the phantom field into matter.
Figure~\ref{fig:evol_densities} shows that $a_\text{eq}$ is shifted to later times for either sign of $\delta$.%
%

Although the phantom field carries negative kinetic energy and has no potential, the potential energy of the quintessence field dominates the dark-energy density at late times, and $\rho_\text{de}=\rho_\phi+\rho_\sigma>0$ at low redshift. 
%
%
%
%

\section{The cosmic microwave background anisotropies} \label{sec:cmban}

\begin{figure}
    \centering
    \includegraphics[width=\linewidth]{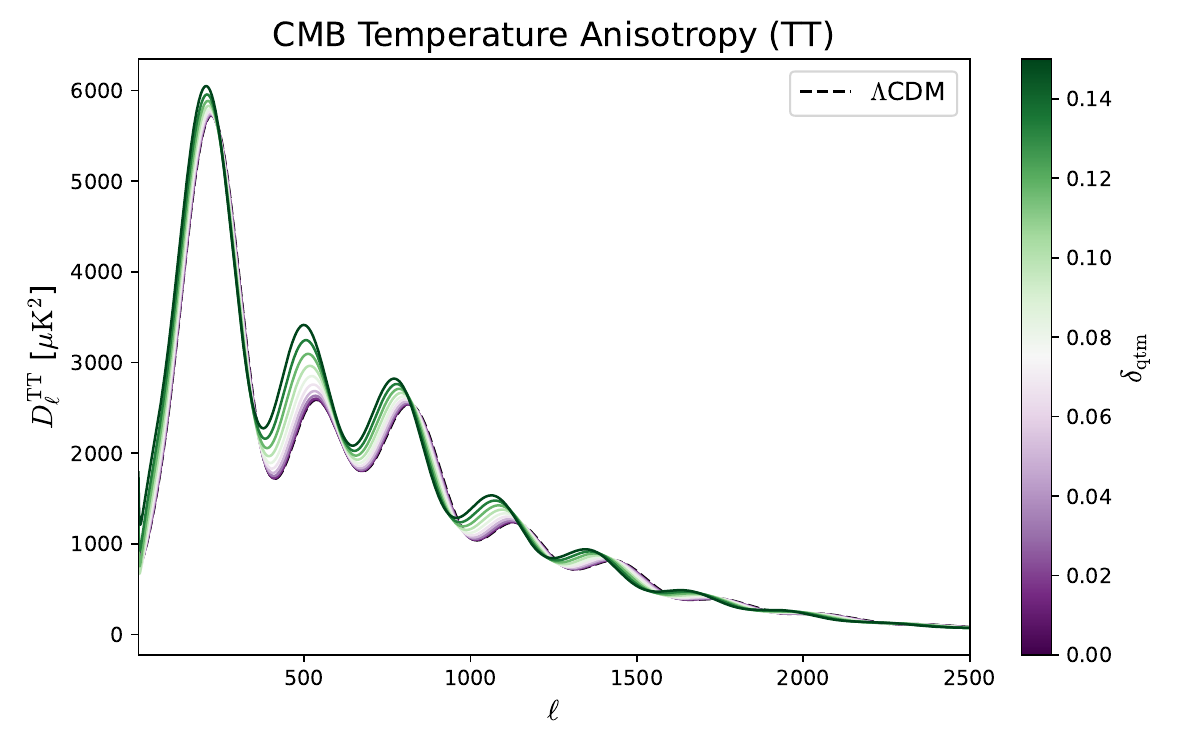}
    \caption{Effect of the coupling $\delta$ on the CMB temperature power spectrum, with background-only calculation.}
    \label{fig:cmb_varying_delta}
\end{figure}

In this section we examine how the modified background expansion alters the CMB temperature power spectrum. The fields and the coupling enter only through $H(z)$ and $\rho_m(a)$, and for the perturbations of photons, baryons and dark matter, we leave them as the standard equations. 
%
%
Figure~\ref{fig:cmb_varying_delta} shows that, relative to $\Lambda$CDM, increasing $|\delta|$ raises the acoustic peaks and shifts them to lower multipoles. 
The coupling transfers energy from the phantom field to both cold dark matter and baryons.
The lower matter density delays matter--radiation equality and leaves more radiation around recombination, so the gravitational potentials decay more strongly.

Before recombination the decay enhances the radiation driving of the acoustic oscillations, and after it the decay enhances the early integrated Sachs--Wolfe effect. Both raise the low-order peaks.
The second peak gains an extra boost because baryons are also coupled.
Their lower density at recombination reduces the baryon loading $R_b=3\rho_b/(4\rho_\gamma)$, which weakens the odd--even peak asymmetry and lifts the second peak relative to the first and third peaks.
%
%

\begin{figure}
    \centering
    \includegraphics[width=\linewidth]{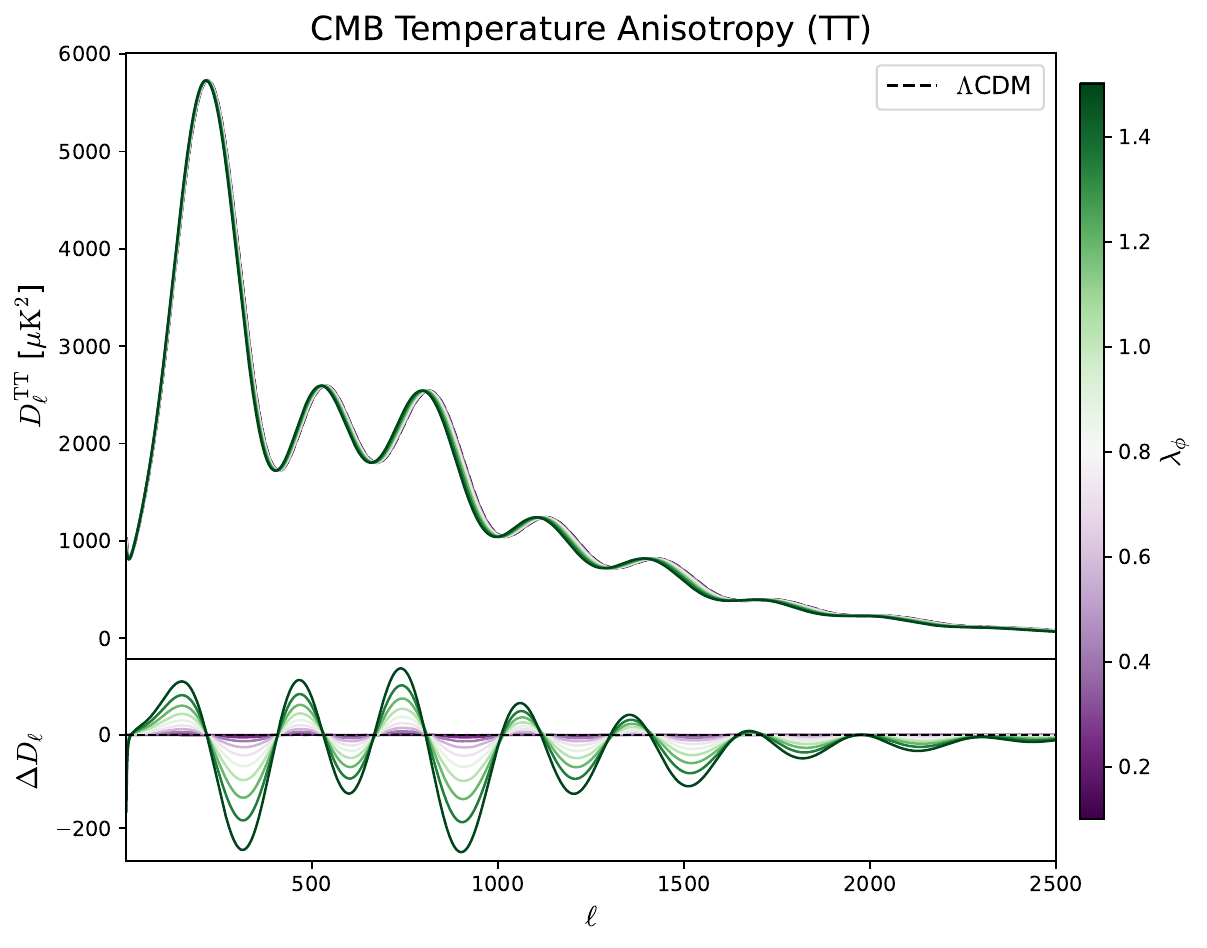}        
    \caption{Effect of the slope $\lambda_\phi$ on the CMB temperature power spectrum}
    \label{fig:cmb_vary_lambda}
\end{figure}

The peak multipoles scale as $\ell\propto\theta_s^{-1}$, where the angular acoustic scale is
\begin{equation}
    \theta_s=\frac{r_{\rm s}(z_*)}{(1+z_*)\,D_A(z_*)}=\frac{r_{\rm s}(z_*)}{D_M(z_*)},
\end{equation}
%
where the comoving sound horizon is given by
\begin{equation}
    r_{\rm s} = \int_{z_*}^{\infty}\frac{c_s}{H(z)}\dd{z},
\end{equation}
with $c_s=c/\sqrt{3(1+R_b)}$, the sound speed of the photon--baryon fluid.
Increasing $|\delta|$ enlarges $r_{\rm s}$, because the lower matter density reduces $H(z)$ before recombination and the lower $R_b$ raises $c_s$. A larger $r_{\rm s}$ moves the peaks to lower $\ell$.
The lower matter density and the negative $\rho_\sigma$ also reduce $H(z)$ after recombination, which increases $D_M(z_*)$ and moves the peaks back toward higher $\ell$. The net shift depends on the balance between the two and on which parameters are held fixed.
The slope $\lambda_\phi$ of the quintessence potential affects only the late-time expansion, because $\Omega_\phi$ is negligible before recombination. For fixed physical densities, it leaves $r_{\rm s}$ unchanged and acts through $D_M(z_*)$ and the late integrated Sachs--Wolfe effect. 
Over the range shown in Fig.~\ref{fig:cmb_vary_lambda}, the change in $D_M(z_*)$ shifts the peak positions, whereas the late integrated Sachs--Wolfe effect alters only the low multipoles.

\section{Observational constraints} \label{sec:obscon}

We use observational data from BAO measurements of distance at certain redshift and luminosity distances from Supernovae Ia to constrain the coupled quintom model. BAO distances are measured in units of the sound horizon $r_{\rm s}$ at the drag epoch redshift $z_{\rm d}$ given by $r_{\rm d}\equiv r_{\rm s}(z_{\rm d})$. There are three kinds of BAO distances; the transverse distance $D_{M}$, the line-of-sight distance $D_{H}$, and the average distance $D_{V}$. They are explicitly
\begin{eqnarray}
D_{M}(z)&=&\frac{c}{H_{0}}\int^{z}_{0}\frac{H_{0}}{H(z')}dz', D_{H}(z)=\frac{c}{H(z)}, \notag\\
D_{V}(z)&=&\left( zD_{M}(z)^{2}D_{H}(z) \right)^{1/3}.
\end{eqnarray}

\begin{itemize}
    \item {\it Baryon Acoustic Oscillations~(BAO)}: The DESI DR2 is used for distance measurements at each redshift as listed in Ref.~\cite{DESI:2025zgx}. This contains $D_{V}/r_{\rm d}$ measurements for redshift in the range $0.1<z<0.4$, BAO distance measurements of $D_{M}/r_{\rm d}$ and $D_{H}/r_{\rm d}$ for $0.4<z<2.1$, and the Ly$\alpha$ measurements in $1.8<z<4.2$, all tracers provided by \verb|Cobaya|~\cite{Torrado:2020dgo}.
\end{itemize}

For the external cosmological datasets, we use:

\begin{itemize}
\item {\it Supernovae Ia~(SN Ia)}: The PantheonPlus dataset~\cite{scolnic2022pantheon+} containing SNe Ia in the redshift range $0.001<z<2.26$.
\item {\it The Cosmic Microwave Background}: 
We use the CMB distance priors of Ref.~\cite{Chen_Huang_Wang_2019}, compressed from the Planck 2018 TT,TE,EE+lowE posterior~\cite{aghanim2020planck}.
These priors were validated for models that modify only the late-time expansion and keep the standard $N_{\rm eff}$~\cite{Chen_Huang_Wang_2019}. In our model the matter density does not scale as $a^{-3}$ and $N_{\rm eff}$ is free. 
The compressed likelihood is therefore only an approximation~\cite{Pedrotti:2026ncd,Zhai:2019nad}, and our CMB-combined results should be read as constraints from the CMB geometry alone. Since the priors carry no CMB-lensing information, the appropriate comparison is with the DESI DR2 preference for $w_0w_a$CDM obtained without CMB lensing, $2.4\sigma$--$2.7\sigma$~\cite{DESI:2025zgx}, rather than $3.1\sigma$.%

We adopt the compressed observational data vector $\{R(z_*), l_A(z_*), \omega_b\}$ where $\omega_b \equiv \Omega_{b}h^{2}$, which encapsulates the geometric information of the CMB acoustic peaks, where $R$ at the redshift of photon decoupling epoch $z_*$ is the CMB-shift parameter, given by
\begin{equation}
    R(z_*) = \frac{\left(1+z_*\right)}{c}\sqrt{\Omega_m H_0^2}\,D_{\mathrm{A}}\left(z_*\right).
\end{equation}
The shift parameter depends on the matter density through $\Omega_mH_0^2\propto\omega_{m,0}$, which in a standard model also fixes the matter density at decoupling. In our model $\rho_m$ is not conserved, and the matter density at decoupling corresponds to the effective value $\omega_m^\text{eff}\equiv\omega_{m,0}\exp[\kappa\delta(\sigma_*-\sigma_0)]\simeq\omega_{m,0}[1-\mathcal{O}(\delta^2)]$. Since $R\propto\sqrt{\omega_m}$, using $\omega_{m,0}$ instead of $\omega_m^\text{eff}$ can overestimates $R$.
We do not correct for this effect and regard it as a limitation of the compressed likelihood.
The acoustic scale, $l_A(z_*)$, defined as 
\begin{equation}
    l_{\mathrm{A}}=\left(1+z_*\right) \frac{\pi D_{\mathrm{A}}(z_*)}{r_s(z_*)}
\end{equation}
where $D_A$ is the physical angular-diameter distance, so that $(1+z_*)D_A=D_M$. $r_s(z_*)$ is the sound horizon distance at $z_*$ defined as 
\begin{equation}
    r_s(z_*)=\int_0^{1 /(1+z_*)} \frac{c\, d a}{a^2 H(a) \sqrt{3\left(1 + a R_b \right)}}.
\end{equation}
%
In our analysis, the detailed recombination history and the value of $z_*$ are computed numerically using
\texttt{HyRec}~\cite{Lee:2020obi,2011PhRvD..83d3513A} within the Boltzmann code \texttt{CLASS}~\cite{2011arXiv1104.2932L,2011JCAP07034B}.


\item {\it Big Bang Nucleosynthesis (BBN) prior}:
In constraining the cosmological parameters of our quintom model, Baryon Acoustic Oscillations (BAO) serve as a critical probe of the universe's expansion history. However, BAO observations act as an uncalibrated standard ruler; they measure the apparent angular size of the sound horizon but yield a degenerate combination of the Hubble parameter and the sound horizon at the drag epoch, $H_0 r_{\rm d}$. To break this parameter degeneracy and isolate the late-time expansion dynamics, an external calibration of $r_{\rm d}$ is required. Following DESI DR2 \cite{tr6y-kpc6}, we adopt a Big Bang Nucleosynthesis (BBN) prior to anchor this scale. The BBN prior—derived from primordial light element abundances (e.g., deuterium)—provides a highly robust, independent measurement of the physical baryon density, $\Omega_b h^2$ \cite{Cooke_Pettini_Steidel_2018,Schöneberg_2026}. Because the baryon-to-photon ratio dictates the speed of sound in the early-universe plasma prior to recombination, this density constraint allows for the precise theoretical calculation of the absolute physical length of $r_{\rm d}$. By calibrating the BAO ruler via BBN, we effectively break the $H_0-r_{\rm d}$ degeneracy, allowing the BAO data to constrain the dynamical dark energy parameters essential to the quintom framework without relying on the full model-dependent assumptions of the cosmic microwave background. Therefore, we use BBN prior as an external dataset, adopting the estimates for $\Omega_b h^2$ and $N_{\rm eff}$ determined by \cite{Schöneberg_2026}. That is,
\begin{equation}
\begin{aligned}
    \Omega_b h^2 &= 0.02196 \pm 0.00063, \\
    N_{\rm eff} &= 2.904 \pm 0.21,
\end{aligned}
\end{equation}
while accounting for the covariance matrix between $\Omega_b h^2$ and $N_{\rm eff}$. We utilize this prior to break the parameter degeneracy and isolate $H_0$, denoting this specific data combination simply as `BBN' throughout our results.
\end{itemize}

\begin{table*}[t]
\centering
\renewcommand{\arraystretch}{1.3} 
\caption{Summary of cosmological parameter constraints from DESI DR2 BAO (labelled as `DESI') combined with external datasets and priors, evaluated within the baseline $\Lambda$CDM and extended Quintom models. To facilitate direct comparison, both marginalized statistical constraints and best-fit values are provided side-by-side for each parameter. The marginalized results (denoted `Marg.') quote the posterior means and 68\% credible intervals, or the 68\% upper limits where only one-sided constraints are derived. The `Best-Fit' columns report the exact parameter values corresponding to the maximum likelihood sample found within the posterior chains for each respective dataset combination.}
\label{tab:merged_cosmo_constraints}
\begin{tabular*}{\linewidth}{@{\extracolsep{\fill}} l c c c c c c c c @{}}
\toprule
 & \multicolumn{2}{c}{$H_0$} & \multicolumn{2}{c}{$\Omega_{\rm m}$} & \multicolumn{2}{c}{$\lambda_{\rm qtm}$} & \multicolumn{2}{c}{$|\delta_{\mathrm{qtm}}|$} \\
\cmidrule(lr){2-3} \cmidrule(lr){4-5} \cmidrule(lr){6-7} \cmidrule(lr){8-9}
\textbf{Model/Dataset} & Marg. & Best-Fit & Marg. & Best-Fit & Marg. & Best-Fit & Marg. & Best-Fit \\
\midrule
\textbf{$\Lambda$CDM} & & & & & & & & \\
CMB & $66.89 \pm 0.61$ & $66.834$ & $0.3232 \pm 0.0087$ & $0.324$ & --- & --- & --- & --- \\
DESI & --- & --- & $0.295^{+0.013}_{-0.016}$ & $0.294$ & --- & --- & --- & --- \\
DESI+BBN & $68.70 \pm 0.80$ & $68.565$ & $0.295 \pm 0.015$ & $0.293$ & --- & --- & --- & --- \\
DESI+BBN+$\theta_*$ & $68.60^{+0.60}_{-0.69}$ & $68.651$ & $0.2936 \pm 0.0075$ & $0.293$ & --- & --- & --- & --- \\
DESI+BBN+SNIa & $ 68.85\pm 0.85$ & $68.844$ & $ 0.311\pm 0.012$ & $0.309$ & ---  & ---  & ---  & ---  \\ 
DESI+BBN+SNIa+SH0ES & $ 70.95\pm 0.66$ & $70.949$ & $ 0.320\pm 0.012$ & $0.317$ & ---  & ---  & ---  & ---  \\ 
DESI+CMB & $68.04 \pm 0.41$ & $68.043$ & $0.3071 \pm 0.0054$ & $0.307$ & --- & --- & --- & --- \\
DESI+CMB+SNIa & $ 67.96\pm 0.40$ & $67.959$ & $ 0.3083\pm 0.0053$ & $0.308$ & ---  & ---  & ---  & ---  \\
DESI+CMB+SNIa+SH0ES & $ 68.78\pm 0.37$ & $68.794$ & $ 0.2978\pm 0.0047$ & $0.298$ & ---  & ---  & ---  & ---  \\ 
\midrule
\textbf{Quintom} & & & & & & & & \\
DESI & --- & --- & $0.309^{+0.016}_{-0.020}$ & $0.302$ & $0.67^{+0.26}_{-0.60}$ & $0.785$ & $0.079^{+0.029}_{-0.074}$ & $7.220 \times 10^{-3}$ \\
DESI+BBN & $61.6^{+6.4}_{-3.2}$ & $66.074$ & $0.309^{+0.017}_{-0.020}$ & $0.296$ & $< 0.841$ & $0.728$ & $0.083^{+0.038}_{-0.075}$ & $7.789\times 10^{-3}$ \\
DESI+CMB & $ 66.1^{+1.5}_{-0.74}$ & $66.995$ & $ 0.3122^{+0.0062}_{-0.012}$ & $0.304$ & $< 0.673$ & $0.020$ & $ 0.0528^{+0.017}_{-0.0091}$ & $5.481\times 10^{-2}$  \\ 
DESI+BBN+SNIa & $ 61.6^{+6.6}_{-3.0}$ & $63.999$ & $ 0.311^{+0.013}_{-0.016}$ & $0.303$ & $ 0.60^{+0.34}_{-0.21}$ & $0.792$ & $ 0.087^{+0.033}_{-0.083}$ & $6.098\times 10^{-2}$ \\ 
DESI+BBN+SNIa+SH0ES & $ 71.86\pm 0.88$ & $71.962$ & $ 0.315\pm 0.012$ & $0.315$ & $< 0.443$ & $0.186$ & $ 0.0151^{+0.0048}_{-0.014}$ & $2.690\times 10^{-3}$ \\
DESI+CMB+SNIa & $ 66.21\pm 0.84$ & $65.780$ & $ 0.3114\pm 0.0064$ & $0.313$ & $ 0.51^{+0.30}_{-0.20}$ & $0.668$ & $ 0.0521^{+0.017}_{-0.0093}$ & $5.811\times 10^{-2}$ \\ 
DESI+CMB+SNIa+SH0ES & $68.49^{+0.46}_{-0.40}$ & $68.736$ & $0.2983 \pm 0.0049$ & $0.298$ & $< 0.274$ & $0.053$ & $0.0170^{+0.0060}_{-0.016}$ & $1.303\times 10^{-4}$ \\
DESI+SNIa+SH0ES & $73.27 \pm 0.97$ & $73.602$ & $0.306 \pm 0.012$ & $0.299$ & $0.59^{+0.33}_{-0.19}$ & $0.826$ & $0.028^{+0.011}_{-0.025}$ & $3.499\times 10^{-3}$ \\
\bottomrule
\end{tabular*}
\end{table*}

\section{Cosmological Inference}. \label{sec:cosinf}

To evaluate the cosmological constraints of the quintom dark energy and standard $\Lambda$CDM models, we mainly utilize the combination of DESI, BBN and CMB distance priors, and SNIa data with and without SH0ES measurement. The parameter space is explored using the cosmological inference code \texttt{Cobaya}~\cite{Torrado_2021} with Metropolis-Hastings MCMC sampling. The theoretical quintom model and $\Lambda$CDM model are computed using interfaces to \texttt{class\_qtm}, which is our modified version from \texttt{class\_public} \cite{Diego_Blas_2011}. For the combination of DESI, BBN, CMB distance priors, and SNIa data (both with and without SH0ES measurements), we sample the standard cosmological parameters $\{H_0, \Omega_{\mathrm{b}}h^2, \Omega_{\mathrm{c}}h^2\}$ and/or $\{H_0, \Omega_{\mathrm{b}}h^2, \Omega_{\mathrm{m}}\}$, which give the same results. For the quintom model, we also sample its dark energy parameters: $\{\lambda_{\rm qtm}, \delta_{\rm qtm}\}$. Prior ranges on all sampled parameters are uniform and chosen to broadly span the physically viable parameter space. 

Since unanchored BAO measurements strictly constrain relative cosmic distance scales, the dataset is fundamentally sensitive to the parameter $hr_d$, where $h = H_0/(100\,\mathrm{km\,s^{-1}\,Mpc^{-1}})$. Consequently, $H_0$ remains unconstrained by the BAO likelihood alone. To accurately report what the data actually measures without our prior limits skewing the results, we extract $hr_d$ and $\Omega_{\rm m}$ (if we already sample $\Omega_{\rm c}h^2$) as derived parameters from the accepted MCMC chains rather than reporting 1D posterior for the individual parameter $H_0$. Moreover, we use absolute magnitude $M$ prior from SH0ES by setting \verb|use_abs_mag: true| when we use SH0ES dataset. Without SH0ES measurement, we set to \verb|false|, which means we marginalize over the value of $M$. 

The convergence criterion for MCMC sampling is that the Gelman-Rubin statistic satisfies $R-1 < 0.01$. Summary statistics for our chains as well as 1D and 2D marginalized posterior plots are obtained with the \verb|GetDist| software package.

To evaluate the statistical preference between the $\Lambda$CDM paradigm and the quintom dark energy model---an extended model with two dark energy parameters, $\lambda_{\rm qtm}$ and $\delta_{\rm qtm}$---we employ the Deviance Information Criterion (DIC). The DIC evaluates the balance of the goodness-of-fit against a penalty for model complexity, evaluating the effective degrees of freedom constrained by the datasets. The fundamental quantity in this analysis is the deviance, defined by:
\begin{equation}
    D(\theta) = -2\ln \mathcal{L}_{\rm data}(\theta),
\end{equation}
where $\theta$ is a model parameter and $\mathcal{L}_{\rm data}(\theta)$ is the joint likelihood function of the data, which is the product of the individual likelihoods from the BAO, CMB distance prior, and SNIa dataset discussed in Section~\ref{sec:obscon}. Because the cosmological datasets utilized assume Gaussian or asymptotically Gaussian errors, yielding $\mathcal{L}(\theta) \propto \exp(-\chi^2/2)$, the deviance is mathematically identical to the $\chi^2$ statistic: $D(\theta) \equiv \chi^2(\theta)$. Note that the deviance calculation strictly isolates the physical fit to the observables, completely excluding any mathematical penalties introduced by theoretical parameter priors.

In its original formulation~\cite{Spiegelhalter_Best_Carlin_Linde_2002}, the complexity penalty, denoted as $p_D$, is defined by $p_D = \overline{D(\theta)} - D(\bar{\theta})$, where $\overline{D(\theta)}$ is the weighted average of the total $\chi^2$ taken over the accepted MCMC samples and $D(\bar{\theta})$ is the total $\chi^2$ evaluated at the posterior mean of the sampled parameters. However, extended dynamical dark energy models often exhibit highly skewed, non-Gaussian posteriors due to complex parameter correlations and physical boundaries, leading to numerical instability when evaluating $D(\bar{\theta})$~\cite{Celeux_Forbes_Robert_Titterington_2006, Trotta_2008}.

To maintain mathematical stability and rigorous complexity penalization in non-Gaussian posteriors, we follow the framework~\cite{Celeux_Forbes_Robert_Titterington_2006, Grandis_Rapetti_Saro_Mohr_Dietrich_2016} and substitute the posterior mean with the maximum a posteriori (MAP) point, which represents the best-fit parameter values of the posterior density distribution. The complexity penalty is thus computed by:
\begin{equation}
    p_D = \overline{D(\theta)} - D_{\rm MAP},
\end{equation}
where $D_{\rm MAP}$ is the total $\chi^2$ evaluated specifically at the best-fit parameter values. To determine these best-fit values and the corresponding $\chi^2$, we use the \verb|minimize| sampler in \verb|Cobaya|, starting from the MAP point of the chains in the MCMC sampling.

The final criterion for a model is then computed as:
\begin{equation}
    {\rm DIC} = \overline{D(\theta)} + p_D.
\end{equation}
Model selection is determined by calculating the difference between the extended and standard paradigms, $\Delta{\rm DIC} = {\rm DIC}_{\rm quintom} - {\rm DIC}_{\rm \Lambda CDM}$. Under this convention, a negative value indicates an information-theoretic preference for the quintom model. As a rule of thumb when comparing models against the $\Lambda$CDM baseline~\cite{Grandis_Rapetti_Saro_Mohr_Dietrich_2016}, an improvement is considered positive for $|\Delta{\rm DIC}| > 2$, strong for $|\Delta{\rm DIC}| > 5$, and decisive for $|\Delta{\rm DIC}| > 10$.

While the DIC analysis penalizes complexity dynamically based on data behavior, an alternative measure of statistical preference for the quintom model over the $\rm \Lambda CDM$ model is the frequentist hypothesis test using $\Delta \chi^2_{\rm MAP}$, which is the difference between MAP best-fit $\chi^2$ values of two models for the particular data combination:
\begin{equation}
    \Delta \chi^2_{\rm MAP} = \chi^2_{\rm MAP,\;quintom} -\chi^2_{\rm MAP,\;\Lambda CDM}.
\end{equation} 
In contrast to DIC, this frequentist approach applies a static penalty based strictly on the addition of new parameters (the degrees of freedom), which in our work correspond to $\lambda_{\rm qtm}$ and $\delta_{\rm qtm}$. Under this framework, the null hypothesis assumes the baseline ${\rm \Lambda CDM}$ model is true, which effectively restricts the extended parameter space by fixing $\lambda_{\rm{qtm}} = 0$ and $\delta_{\rm{qtm}} = 0$. Rejecting this null hypothesis indicates that the quintom model is statistically preferred. Because the models are nested, Wilks' theorem~\cite{wilks1938large} dictates that the $\Delta\chi^2_{\rm MAP}$ test statistic follows a $\chi^2$ distribution with two degrees of freedom, denoted as $f(x;\nu=2)$. We first compute the $p$-value by evaluating the tail probability:
\begin{equation}
    p = \int_{|\Delta\chi^2_{\rm{MAP}}|}^{\infty} f(x;\nu=2)\;\mathrm{d}x.
\end{equation}
This integral represents the probability of observing a $\Delta \chi^2_{\rm MAP}$ this large or larger, assuming the baseline $\Lambda$CDM model is the true model. By equating this $p$-value to a significance level $\alpha$, we define the minimum threshold required to reject the null hypothesis (where rejection occurs when $p < \alpha$). This formulation also defines a corresponding confidence level, $1-\alpha$, for the statistical preference for the quintom model. To express this evaluation in familiar terms, we translate this confidence level into a symmetric, two-sided Gaussian interval bounded by $S$ standard deviations, $1-\alpha = P(|Z|\leq S)$, where $Z$ follows the standard normal distribution. Finally, $S\sigma$ is defined as the frequentist significance of the parameter extension, and is computed as:
\begin{align}
    S = {\rm CDF}_{\mathcal{N}}^{-1}\left(1-\frac{p}{2}\right),
\end{align}
where $\rm{CDF}_{\mathcal{N}}^{-1}$ is the inverse cumulative distribution function of the standard Gaussian distribution. Note that this framework applies to all data outcomes, but it only yields a non-zero significance ($S > 0\sigma$) when $\Delta\chi^2_{\rm{MAP}} < 0$ under our convention (i.e., when the quintom model provides a better fit to the data). When $\Delta\chi^2_{\rm MAP} \ge 0$, the test concludes that there is no evidence to reject the null hypothesis and $S=0.00\sigma$.

\section{Constraints on Cosmological Parameters}. \label{sec:constr}

\subsection{$\Lambda {\rm CDM}$ Model}

The marginalized and best-fit values of the $\Lambda$CDM model are listed in Table~\ref{tab:merged_cosmo_constraints}. For references, we run the constraints on the combination of DESI+BBN, DESI+BBN+$\theta_{*}$ and DESI+CMB, using BBN $\{\Omega_b h^2,N_{\rm eff}\}$ and CMB distance priors. The marginalized values of $H_{0}$ from DESI+BNN, DESI+CMB are $68.70\pm0.80~(68.04\pm 0.41)$~km s$^{-1}$ Mpc$^{-1}$ with the corresponding matter density $\Omega_{m}=0.295\pm0.015~(0.3071\pm 0.0054)$ respectively. These values of Hubble parameter are slightly higher than $H_{0}=66.89\pm0.61$~km s$^{-1}$ Mpc$^{-1}$ from the CMB only. In contrast, the matter density value from only CMB prior is subsequently larger $\Omega_{m}=0.3232\pm 0.0087$. 

\begin{figure*}
    \centering
    \includegraphics[width=0.49\linewidth]{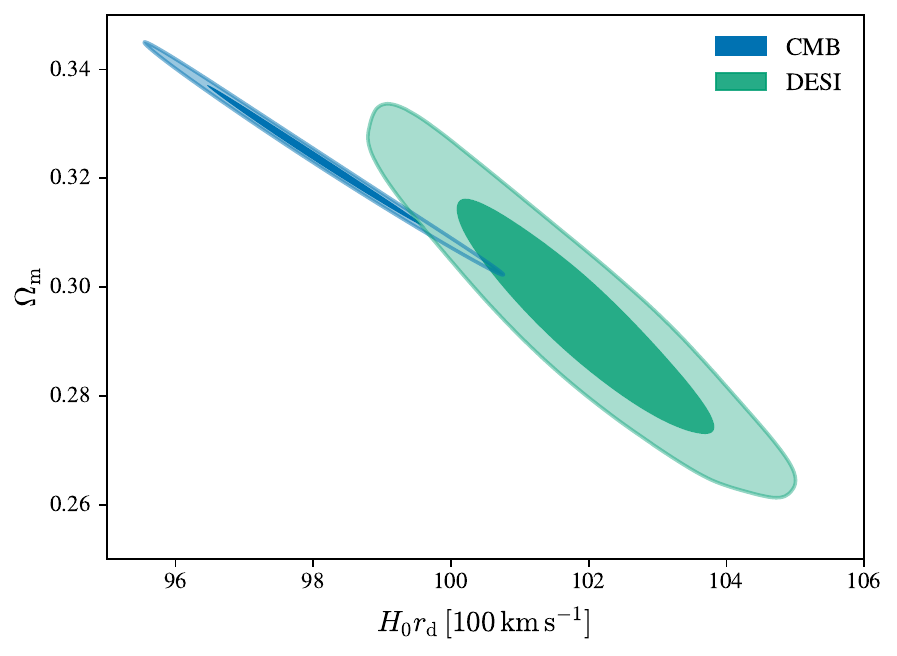}
    \includegraphics[width=0.49\linewidth]{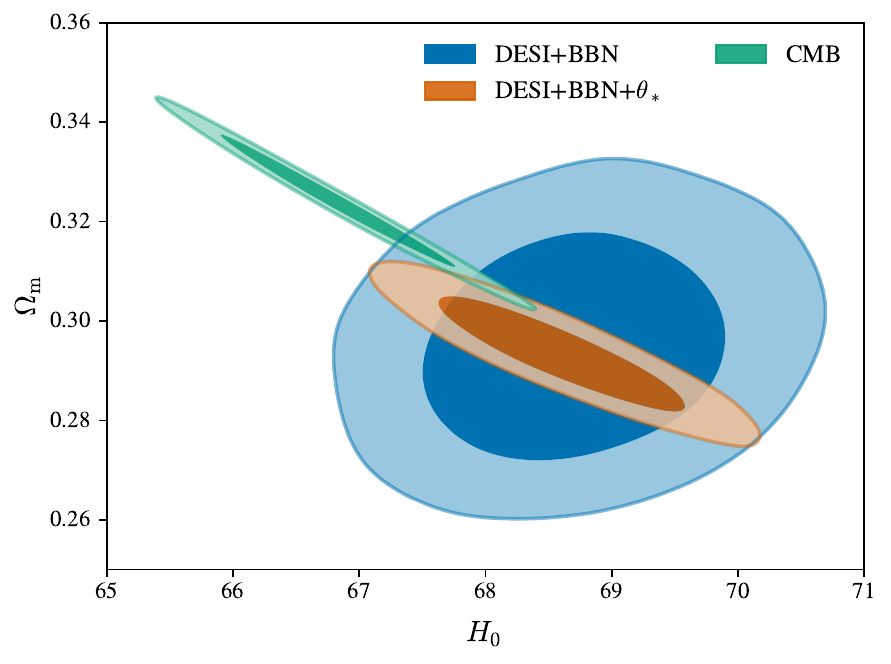}
    \caption{2D marginalized posterior distributions for the flat $\Lambda$CDM model, showing 68\% and 95\% confidence intervals derived from our independent MCMC sampling. \textit{Left}: Constraints in the $H_0 r_{\rm d} - \Omega_{\rm m}$ parameter space. This panel illustrates the intrinsic geometric degeneracy of the uncalibrated DESI BAO measurements (green), which strictly constrain the product $H_0 r_{\rm d}$ rather than either parameter independently. The posterior derived from the CMB distance prior (blue) is overlaid to demonstrate its near-orthogonality and complementarity to the late-time BAO data. \textit{Right}: Constraints in the $H_0 - \Omega_{\rm m}$ plane, demonstrating the impact of breaking the geometric degeneracy. The broad baseline constraints from the combined DESI and BBN likelihoods (blue) are significantly tightened upon the inclusion of the gaussian prior, $\theta_*$ (orange). This prior effectively calibrates the sound horizon, yielding restricted posteriors that approach the constraining power and location of the constraints derived from the CMB distance prior (green)}
    \label{fig:placeholder}
\end{figure*}

\subsection{Quintom Model and Comparison to $\Lambda {\rm CDM}$}

Table \ref{tab:merged_cosmo_constraints} shows the marginalized and best-fit values of interacting quintom model for the following combination of datasets, \{DESI, SNIa and SH0ES\} and priors, \{BBN, CMB\}. The low-redshift datasets DESI+BBN yields relatively small values of the {\it best-fit} $H_{0}=66.074$~km s$^{-1}$ Mpc$^{-1}$ with $\Omega_{m}=0.296$ and $\{\lambda_{\rm qtm},|\delta_{\rm qtm}|\}=(0.728,0.008)$, preferring the small phantom-matter coupling. The $H_{0}$ value is slightly smaller than the corresponding $\Lambda$CDM value with almost the same $\Omega_{m}$.  

Adding SNIa, DESI+BBN+SNIa yields even smaller best-fit value $H_{0}=63.999$~km s$^{-1}$ Mpc$^{-1}$ with $\Omega_{m}=0.303$ and $\{\lambda_{\rm qtm},|\delta_{\rm qtm}|\}=(0.792,0.061)$. Due to the high value of absolute magnitude $M$ in SH0ES, DESI+BBN+SNIa+SH0ES gives the musch higher best-fit value $H_{0}=71.962$~km s$^{-1}$ Mpc$^{-1}$ with $\Omega_{m}=0.315$ and $\{\lambda_{\rm qtm},|\delta_{\rm qtm}|\}=(0.186,0.003)$, still preferring a very small phantom-matter coupling. Notably, without BBN prior, DESI+SNIa+SH0ES gives $H_{0}=73.602$~km s$^{-1}$ Mpc$^{-1}$ with larger $\Omega_{m}=0.299$ and $\{\lambda_{\rm qtm},|\delta_{\rm qtm}|\}=(0.826,0.003)$, an almost non-interacting quintom. 

For combination of low and high redshift datasets, DESI+CMB gives the best-fit value $H_{0}=66.995$~km s$^{-1}$ Mpc$^{-1}$ with $\Omega_{m}=0.304$ and $\{\lambda_{\rm qtm},|\delta_{\rm qtm}|\}=(0.020,0.055)$. These $\{H_{0},\Omega_{m}\}$ values are slightly smaller than the corresponding $\Lambda$CDM values. DESI+CMB+SNIa gives even smaller $H_{0}=65.780$~km s$^{-1}$ Mpc$^{-1}$ with larger $\Omega_{m}=0.313$ and $\{\lambda_{\rm qtm},|\delta_{\rm qtm}|\}=(0.668,0.058)$. On the other hand, DESI+CMB+SNIa+SH0ES remarkably prefers {\it non-interacting} quintom with the best-fit values $H_{0}=68.736$~km s$^{-1}$ Mpc$^{-1}$ with $\Omega_{m}=0.298$ and $\{\lambda_{\rm qtm},|\delta_{\rm qtm}|\}=(0.053,0.0001)$. The absolute magnitude value used in SH0ES data sampling, $M_{b}=-19.40$, constrains the parameter space of quintom model to the region where phantom-matter couping is extremely weak. For CMB-combined datasets in Figure~\ref{fig:triangle_plot_quintom_CMB} which marginalized absolute magnitude, the marginalized value is $M_{b}=-19.48$.

For comparison of constraints on notable cosmological parameters, Figure~\ref{fig:triangle_plot_quintom_CMB} and \ref{fig:triangle_plot_quintom_BBN} show the triangular plots of interacting quintom model subject to various combination of datasets. Figure~\ref{fig:lcdm_quintom_comparison_1} compares $\Lambda$CDM and quintom model for the DESI+CMB+SNIa datasets. Notably, interacting quintom model prefers lower $H_{0}$ and $\sigma_{8}$ comparing to the $\Lambda$CDM. Moreover, the interacting quintom model prefers a slightly lower $\Omega_{\rm b}h^{2}$ but higher $\Omega_{\rm m}$ than the $\Lambda$CDM.

The DIC comparisons between the quintom and $\Lambda$CDM model are summarized in Table~\ref{tab:dic_analysis}. For combined datasets except DESI only, the interacting quintom model fits observational data better than the $\Lambda$CDM. 

\begin{table}
    \centering
    \begin{tabular}{l  c  c  c}
    \toprule
         \textbf{Dataset} & $\Delta\chi^2_{\rm MAP}$ & $S\sigma$ & $\Delta$DIC   \\
         \hline
            DESI+CMB& $-6.51$& $2.07\sigma$& $-3.16$\\
            DESI+CMB+SNIa& $-7.69$& $2.30\sigma$& $-3.39$\\
    \bottomrule
    \end{tabular}
    \caption{Statistical comparison of the quintom model against the baseline $\Lambda$CDM model across various observational data combinations. The table reports the difference in $\chi^2_{\rm MAP}$ values ($\Delta\chi^2_{\rm{MAP}}$), the resulting frequentist significance of the parameter extension ($S\sigma$), and the difference in the Deviance Information Criterion ($\Delta$DIC). Negative values for $\Delta\chi^2_{\text{MAP}}$ indicate a better raw fit for the quintom model, and negative $\Delta$DIC values denote a preference for the quintom model.}
    \label{tab:dic_analysis}
\end{table}

\begin{table}[]
    \centering
    \begin{tabular}{l c}
        \toprule
        \textbf{Model/Dataset} & $\sigma_8$ \\
        \hline
        $\mathbf{\Lambda {\rm CDM}}$ &  \\
        CMB & $ 0.8165\pm 0.0047$ \\ 
        DESI & $ 0.800^{+0.051}_{-0.062}$ \\ 
        DESI+BBN & $ 0.800\pm 0.037$ \\ 
        DESI+BBN+$\theta_*$ & $ 0.7954^{+0.0057}_{-0.0051}$ \\
        DESI+BBN+SNIa & $ 0.835\pm 0.031$ \\ 
        DESI+BBN+SNIa+SH0ES & $ 0.890\pm 0.030$ \\ 
        DESI+CMB & $ 0.8077\pm 0.0033$ \\
        DESI+CMB+SNIa & $ 0.8084\pm 0.0032$ \\ 
        DESI+CMB+SNIa+SH0ES & $ 0.8022\pm 0.0030$ \\ 
        \hline
        \textbf{Quintom} &  \\ 
        DESI & $ 0.623^{+0.19}_{-0.091}$ \\ 
        DESI+BBN & $ 0.606^{+0.20}_{-0.089}$ \\ 
        DESI+BBN+SNIa & $ 0.607^{+0.21}_{-0.094}$ \\
        DESI+BBN+SNIa+SH0ES & $ 0.863\pm 0.033$ \\ 
        DESI+CMB & $ 0.739\pm 0.030$ \\ 
        DESI+CMB+SNIa & $ 0.742\pm 0.028$ \\ 
        DESI+CMB+SNIa+SH0ES & $ 0.791^{+0.013}_{-0.0042}$ \\ 
        DESI+SNIa+SH0ES & $ 0.813\pm 0.045$ \\ 
        \hline
    \end{tabular}
    \caption{Values of $\sigma_{8}$ parameter in $\Lambda$CDM and quintom models for all combined datasets. Quintom model gives generally smaller values for datasets without SH0ES.}
    \label{tab:placeholder}
\end{table}

\begin{figure}
    \centering
    \includegraphics[width=\linewidth]{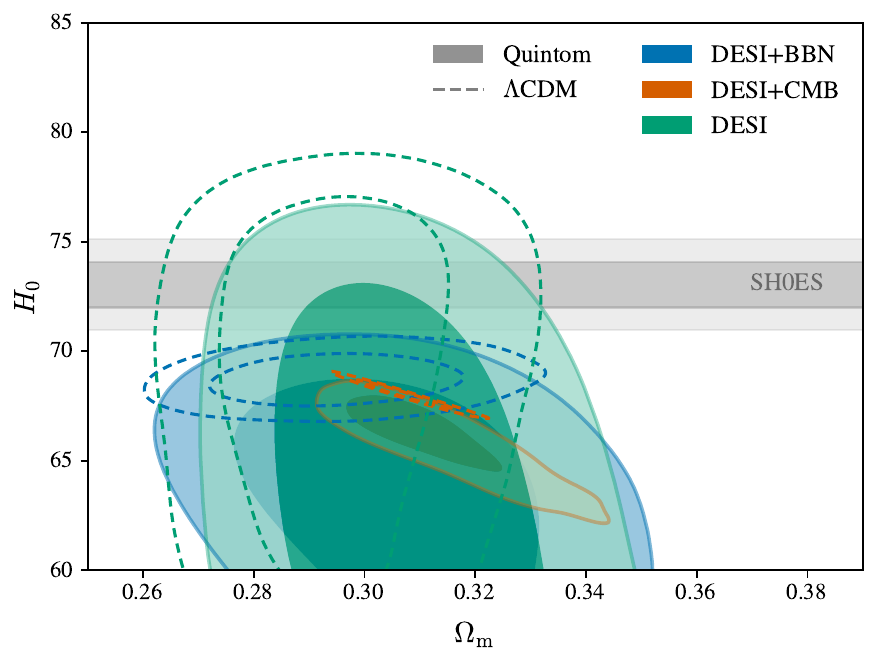}
    \caption{2D marginalized posteriors (68\% and 95\% credible intervals) in the $H_0 - \Omega_{\rm m}$ plane. Filled and dashed contours represent the Quintom and $\Lambda$CDM models, respectively, evaluated using DESI (teal), DESI+BBN (blue), and DESI+CMB (orange) data. The horizontal gray band indicates the SH0ES local $H_0$ measurement.}
    \label{fig:5}
\end{figure}

\begin{figure}
    \centering
    \includegraphics[width=\linewidth]{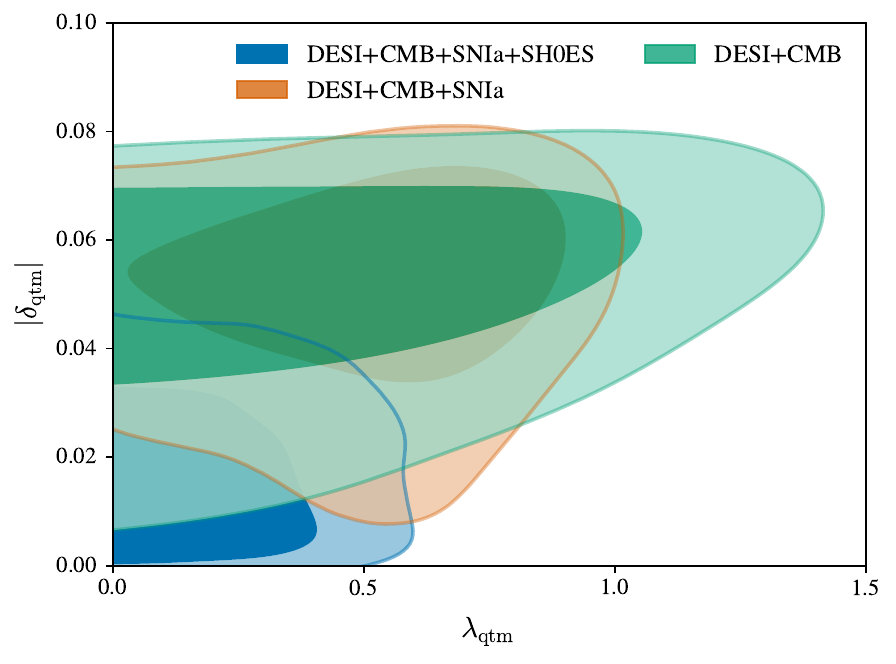}
    \caption{2D marginalized posteriors (68\% and 95\% credible intervals) for the Quintom parameters $\lambda_{\rm qtm}$ and $|\delta_{\rm qtm}|$. Including late-time SNIa data (peach) breaks parameter degeneracies present in the baseline DESI+CMB constraints (teal). Furthermore, incorporating the local $H_0$ prior (SH0ES, dark blue) severely restricts the available parameter space, driving both dynamical dark energy parameters closer to zero.}
    \label{fig:6}
\end{figure}

\subsection{$H_{0}$ and $\sigma_{8}$ tension}

Lastly, we address Hubble and $\sigma_{8}$ tension in the interacting quintom model. From Table~\ref{tab:merged_cosmo_constraints}, DESI+CMB, DESI+CMB+SNIa and DESI+CMB+SH0ES datasets, all three combinations of low-redshift and high-redshift CMB data give a consistent $H_{0}=67.0-68.7$~km s$^{-1}$ Mpc$^{-1}$. However, using only low-redshift datasets, DESI+SNIa+SH0ES gives the best-fit value $H_{0}=73.62$~km s$^{-1}$ Mpc$^{-1}$, the Hubble tension originated from SH0ES remains in this model. This is due to the absolute magnitude prior of SH0ES. 

On the other hand from Table~\ref{tab:placeholder}, DESI+CMB and DESI+CMB+SNIa datasets give the value of $\sigma_{8}$ parameter in the range $\sigma_{8}=0.74$, consistent with the value $\sigma_{8}=0.753\pm 0.011$ from low redshift galaxy clustering and galaxy-galaxy lensing~\cite{DES:2021bpo}. With SH0ES dataset, $\sigma_{8}$ increases to about $0.79-0.81$. There is no distinctive $\sigma_{8}$ tension in the interacting quintom model when SH0ES is excluded.  

\section{Local scale constraints on the phantom-matter coupling}  \label{SectCassini}
The phantom-matter coupling in Eqn.~\eqref{eq:phm} generates a fifth force on the matter as will be shown in the following. Consider the static pressureless matter with $T^{\mu}_{\nu,m}=\rho_{m}u^{\mu}u_{\nu}, T_{m}=-\rho_{m}$~(note the dependence of sign on the signature of spacetime metric) where $u^{\mu}$ is the 4-velocity. Eqn.~\eqref{eq:phm} gives
\begin{equation}
u_\nu \nabla_\mu (\rho_M u^\mu) + \rho_M u^\mu \nabla_\mu u_\nu = -\kappa \delta \rho_M \partial_\nu \sigma  \label{eq:temp}
\end{equation}
Contracting this equation with $u^\nu$ (applying the identities $u^\nu u_\nu = -1$ and $u^\nu \nabla_\mu u_\nu = 0$) isolates the modified continuity equation:
\begin{equation}
\nabla_\mu (\rho_M u^\mu) = \kappa \delta \rho_M u^\mu \partial_\mu \sigma.
\end{equation}
Substituting back into \eqref{eq:temp} to obtain mass-independent equation of motion for a matter particle,
\begin{equation}
u^\mu \nabla_\mu u_\nu = -\kappa \delta (\partial_\nu \sigma + u_\nu u^\mu \partial_\mu \sigma).
\end{equation}
In the weak-field, non-relativistic limit ($u^0 \approx 1$, $u^i = v^i \ll 1$) and assuming a static field configuration ($\partial_0 \sigma = 0$), the spatial components ($\nu = i$) simplify to
\begin{equation}
\frac{d^2 x^i}{d\tau^2} + \Gamma^i_{00} = -\kappa \delta \partial_i \sigma.
\end{equation}
Identifying $\Gamma^i_{00} = -\partial_i \Phi_N$ and $\Phi_{N}=GM/r$ as the standard Newtonian gravitational gradient, the total acceleration vector $\vec{a}$ is expressed as
\begin{equation}
\vec{a} = \vec{\nabla}\Phi_N - \kappa \delta \vec{\nabla}\sigma.
\end{equation}
The second term defines the explicit fifth force per unit mass
\begin{equation}
\vec{a}_5 = - \kappa \delta \vec{\nabla}\sigma.  \label{eq:a5}
\end{equation}

Next we obtain a phantom profile by varying the action with respect to $\sigma$. Eqn.~\eqref{eq:phm} establishes how local matter sources the scalar field profile.
\begin{equation}
S_{\sigma}+S_{m}= \int~d^{4}x \sqrt{-g}\left(\frac{1}{2}\nabla_{\mu}\sigma\nabla^{\mu}\sigma+\mathcal{L}_{m}+\mathcal{L}_{m\sigma}\right),
\end{equation}
where $\mathcal{L}_{m}$ is the matter Lagrangian density and $\mathcal{L}_{m\sigma}=\kappa\delta T_{m}\sigma$ is the conformal phantom-matter coupling term. The equation of motion of the phantom is then given by
\begin{equation}
\nabla_\mu \nabla^\mu \sigma = -\kappa \delta \rho_M.
\end{equation}
To obtain the local spatial field profile, we place a static localized point mass $M$ at the origin, described by the density distribution $\rho_M(\vec{r}) = M \delta^3(\vec{r})$. The Poisson-like equation becomes
\begin{equation}
\nabla^2 \sigma = -\kappa \delta M \delta^3(\vec{r}).
\end{equation}
Utilizing the standard Green's function for the spatial Laplacian, $\nabla^2 \left(\frac{1}{4\pi r}\right) = -\delta^3(\vec{r})$, the static field profile surrounding the mass can be obtained,
\begin{equation}
\sigma(r) = \frac{\kappa \delta M}{4\pi r}  \label{eq:sigr}.
\end{equation}
Substitute the phantom profile \eqref{eq:sigr} to the fifth-force acceleration \eqref{eq:a5} to get
\begin{equation}
\vec{a}_5 =\frac{\kappa^2 \delta^2 M}{4\pi r^2} \hat{r},
\end{equation}
regardless of the sign of $\delta$, the acceleration is always radially outward~(universally repulsive).
Since $\kappa^2 = 8\pi G$, the localized fifth-force acceleration simplifies to:
\begin{equation}
\vec{a}_5 = \frac{2 \delta^2 G M}{r^2} \hat{r}.
\end{equation}
The conformal phantom-matter coupling $\mathcal{L}_{m\sigma}=\kappa\delta T_{m}\sigma$ produces a universally repulsive interaction on baryonic matter and dark matter. 

To compare with Cassini constraints~\cite{Bertotti:2003rm}, we cast the effects of phantom-matter fifth force into the effective metric form, this can be done since the phantom-matter coupling is assumed gravitational and universal. First, we express the weak-field parametrized post-Newtonian~(PPN) metric surrounding a spherically symmetric mass as:
\begin{align}
g_{00} &= -(1 - 2\Phi_{\text{obs}}), ~g_{ij} = \delta_{ij}(1 + 2\gamma\Phi_{\text{obs}}) \label{eq:cass}
\end{align}
where $\Phi_{\text{obs}}$ is the total effective gravitational potential measured via local orbital acceleration. Combining standard Newtonian gravity with our derived fifth-force acceleration yields the net acceleration vector:
\begin{equation}
\vec{a}_{\text{total}} = -\frac{GM}{r^2}(1 - 2\delta^2)\hat{r} \implies \Phi_{\text{obs}} = \Phi_N(1 -2\delta^2),
\end{equation}
where $\Phi_N = \displaystyle{\frac{GM}{r}}$.
Because the conformal metric mapping shifts spatial curvature via the scalar field backreaction, the effective metric becomes~(see Appendix~\ref{appb})
\begin{align}
\tilde{g}_{00} &= (1+2\kappa\delta\sigma)g_{00}\simeq -\left( 1-2\Phi_{\rm N}(1-2\delta^2)\right), \\
\tilde{g}_{ij} &= \delta_{ij}\left(1 + 2\Phi_{\rm N}(1+2\delta^2)\right),  \label{eq:geff}
\end{align}
where we have used \eqref{eq:sigr}. Comparing with the PPN metric in \eqref{eq:cass}, the ratio of these expressions allows us to isolate the PPN parameter $\gamma$:
\begin{equation}
\gamma = \frac{1 + 2\delta^2}{1 - 2\delta^2}.
\end{equation}
To compare this with solar system bounds, we calculate the absolute deviation from the General Relativity baseline ($\gamma_{\text{GR}} = 1$):
\begin{equation}
|\gamma - 1| =   \frac{4\delta^2}{1 - 2\delta^2} 
\end{equation}
The Cassini tracking experiment constrains this metric deviation to
\begin{equation}
|\gamma - 1| \le 2.1\pm 2.3 \times 10^{-5} \implies |\delta| < 3.3 \times 10^{-3},  \label{eq:delb}
\end{equation}
using the upper bound on $|\gamma-1|$. Interestingly, this solar-system bound is still consistent with the best-fit values from cosmological observational combined datasets with SH0ES, i.e., DESI+BBN/CMB+SNIa+SH0ES and DESI+SNIa+SH0ES in Table~\ref{tab:merged_cosmo_constraints}.

\section{Conclusions and Discussion} \label{sec:con}
We analyzed the quintom model with phantom--matter interaction and apply constraints from DESI, SNIa dataset and BBN/CMB likelihood to the model. The coupled quintom model exhibits phantom crossing behavior at late times and serves as a viable two-component interacting dark energy consistent with current observation, both low and high redshift datasets. Notably, the redshift dependence of $w_{\rm qtm}(z)$ shown in Fig.~\ref{fig:evol_densities} is distinctive from $w_{0}-w_{a}$ model and has a unique characteristic. 

Due to the phantom-matter coupling, quintom model allows a wider range of matter density values consistent with current DESI BAO, SNIa and CMB data. Without absolute magnitude prior, e.g. from SH0ES, interacting quintom model generically prefers lower values of $H_{0}$ and $\sigma_{8}$, slightly lower $\Omega_{\rm b}h^{2}$ and higher $\Omega_{\rm m}$ than $\Lambda$CDM model. 

For the combination of low and high redshift data; DESI+CMB, DESI+CMB+SNIa and DESI+CMB+SNIa+SH0ES all give $H_{0}=67.0-68.7$~km s$^{-1}$ Mpc$^{-1}$ and $\sigma_{8}=0.74-0.79$, consistent with the value $\sigma_{8}=0.753\pm0.011$ from low redshift galaxy clustering and galaxy-galaxy lensing. While the Hubble tension between the datasets with CMB and SH0ES-only dataset remains, there is no $\sigma_{8}$ tension in the interacting quintom model. \\ 

\begin{acknowledgments}
    CK is supported by Fundamental Fund grant from Thailand Science Research and Innovation (TSRI).
\end{acknowledgments}

\appendix
\section{Triangular plots of combined dataset constraints}
\begin{figure*}
    \centering
    \includegraphics[width=\linewidth]{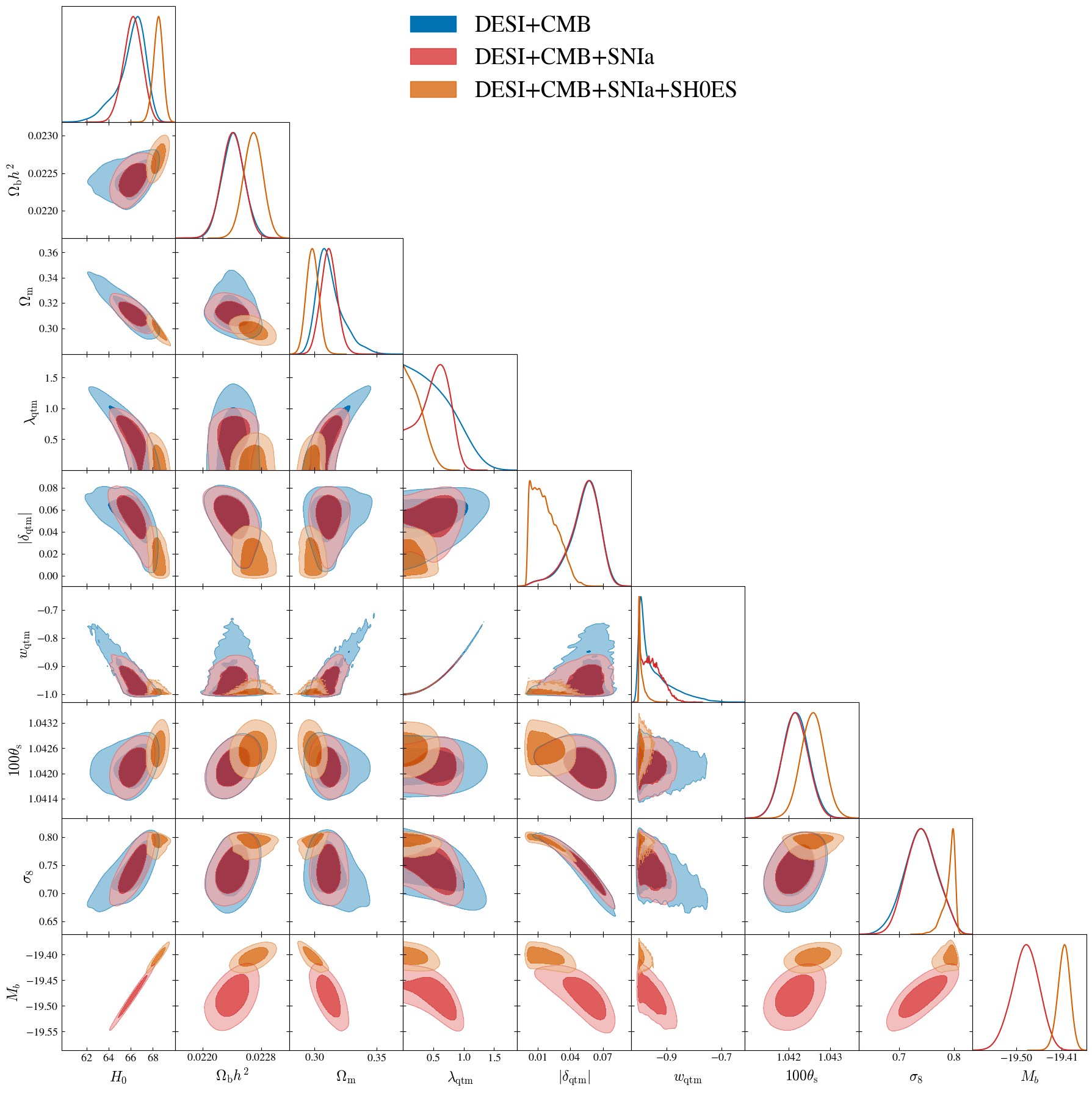}
    \caption{Marginalized 1D posterior distributions and 2D contour plots (68\% and 95\% confidence levels) for the cosmological parameter space $\{H_0, \Omega_{\rm b} h^2, \Omega_{\rm m}, \lambda_{\rm qtm}, \vert{}\delta_{\rm qtm}\vert{}, w_{\rm qtm}, 100\theta_{\rm s}, \sigma_8, M_b\}$ of the Quintom dark energy model constrained by the joint datasets of DESI+CMB (blue), DESI+CMB+SNIa (red), and DESI+CMB+SNIa+SH0ES (orange).}
    \label{fig:triangle_plot_quintom_CMB}
\end{figure*}

\begin{figure*}
    \centering
    \includegraphics[width=1\linewidth]{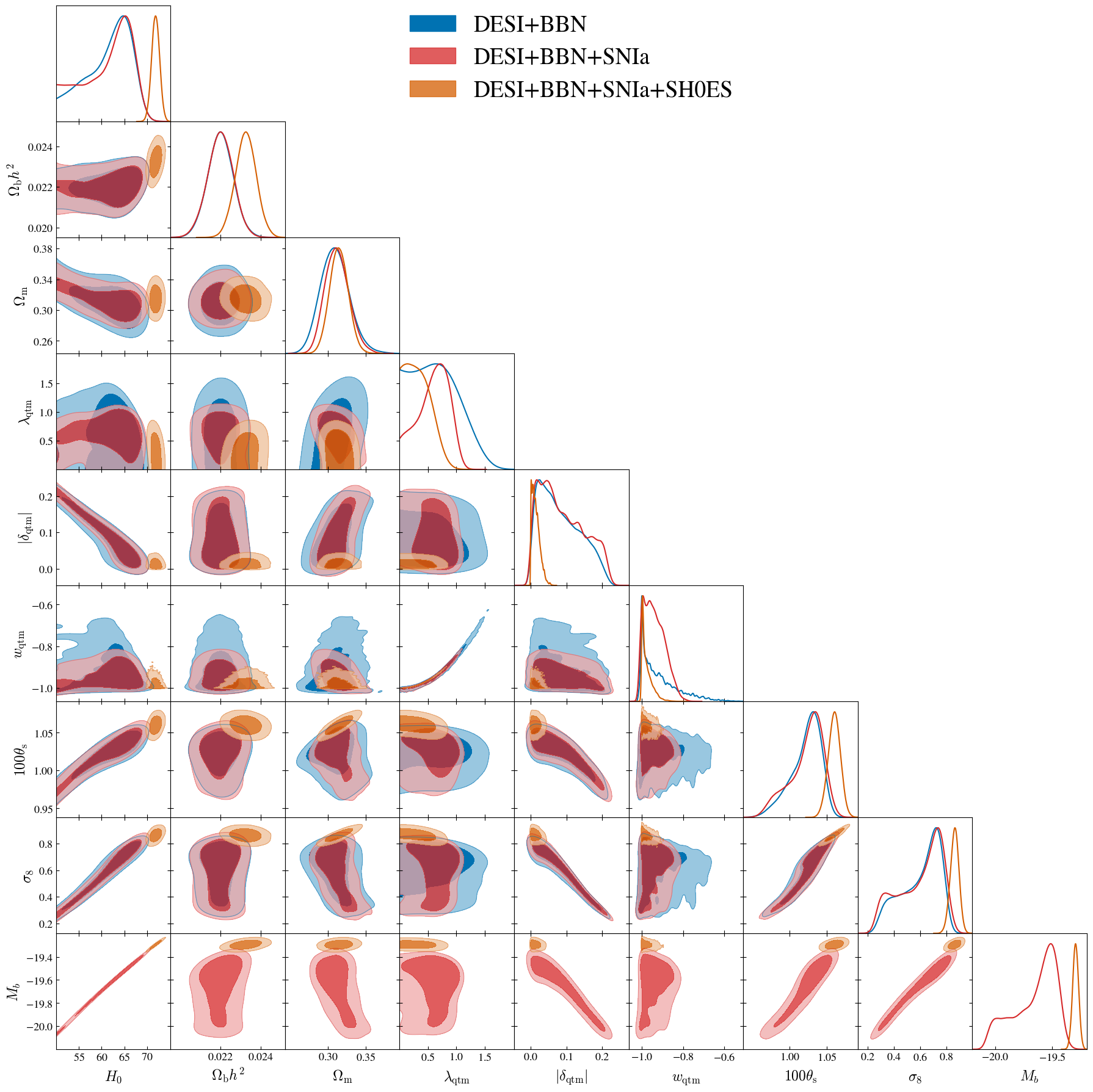}
    \caption{Marginalized 1D posterior distributions and 2D contour plots (68\% and 95\% confidence levels) for the cosmological parameter space $\{H_0, \Omega_{\rm b} h^2, \Omega_{\rm m}, \lambda_{\rm qtm}, \vert{}\delta_{\rm qtm}\vert{}, w_{\rm qtm}, 100\theta_{\rm s}, \sigma_8, M_b\}$ of the Quintom dark energy model constrained by the joint datasets of DESI+BBN (blue), DESI+BBN+SNIa (red), and DESI+BBN+SNIa+SH0ES (orange).}
    \label{fig:triangle_plot_quintom_BBN}
\end{figure*}

\begin{figure*}
    \centering
    \includegraphics[width=\linewidth]{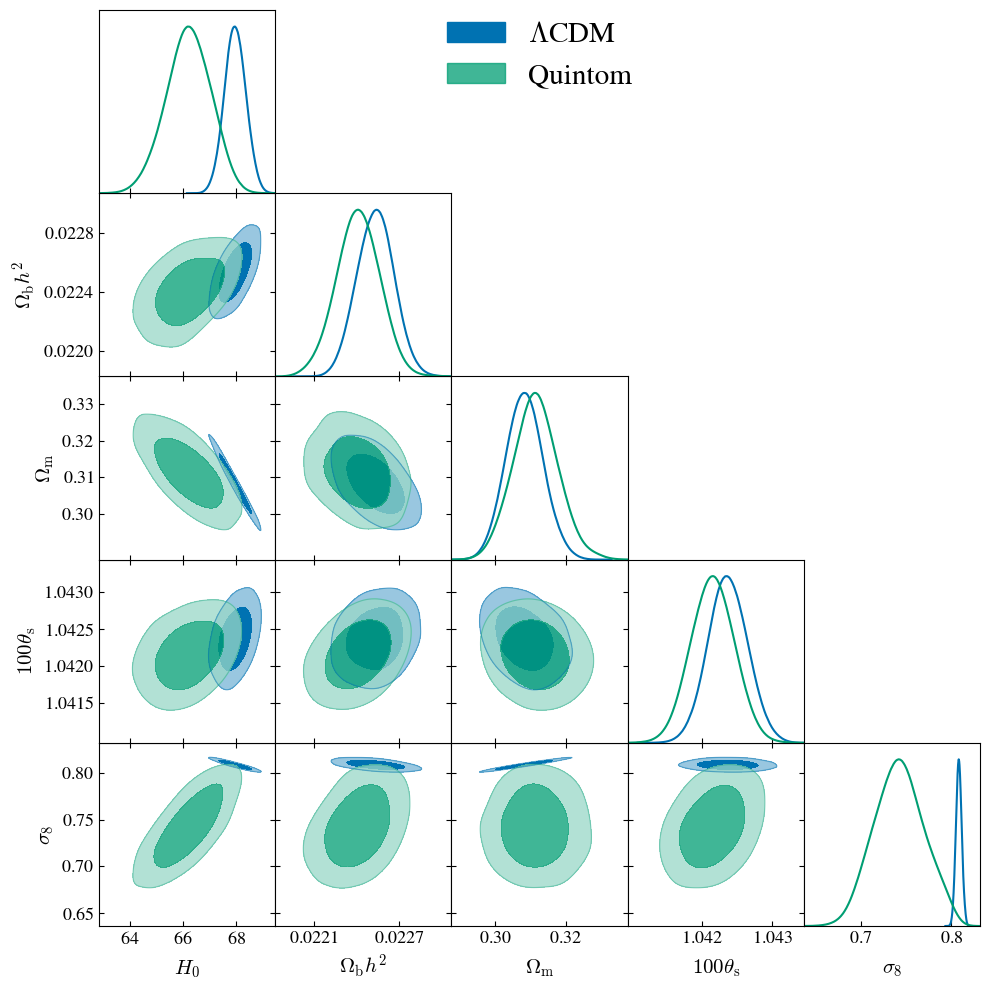}
    \caption{Marginalized 1D posterior distributions and 2D contour plots (68\% and 95\% confidence levels) for the cosmological parameter space $\{H_0, \Omega_{\rm b} h^2, \Omega_{\rm m}, 100\theta_{\rm s}, \sigma_8\}$ comparing the $\Lambda$CDM and Quintom models, constrained by the joint datasets of DESI data, CMB distance priors, and SNIa data.}
    \label{fig:lcdm_quintom_comparison_1}
\end{figure*}

The triangular plots Fig.~\ref{fig:triangle_plot_quintom_CMB}-Fig.~\ref{fig:lcdm_quintom_comparison_1} for all dataset constraints on interacting quintom model and the comparison to the $\Lambda$CDM for DESI+CMB+SNIa datasets are presented in this section.

\section{Derivation of effective metric including phantom-matter coupling effects} \label{appb}

The interaction between non-relativistic matter and the phantom field $\sigma$ is defined via the energy-momentum tensor non-conservation equation:
\begin{equation}
\nabla_\mu T^{\mu}_{\nu(m)} = \kappa \delta T_m \nabla_\nu \sigma \label{eq:apm}
\end{equation}
In scalar-tensor theories, this specific baryonic coupling implies that the physical Jordan-frame metric $\tilde{g}_{\mu\nu}$ (governing matter trajectories and photon paths) is related to the Einstein-frame metric $g_{\mu\nu}$ via a conformal scaling factor $\Omega(\sigma)$:
\begin{equation}
\tilde{g}_{\mu\nu} = \Omega^2(\sigma) g_{\mu\nu},
\end{equation}
leading to the variation
\begin{equation}
\delta \tilde{g}_{\mu\nu} = \Omega^2 \left( \delta g_{\mu\nu} + 2 \frac{d\ln \Omega}{d\sigma} g_{\mu\nu} \delta\sigma \right).
\end{equation}
For the Einstein-frame action
\begin{equation}
S_{m}=\int d^{4}x\sqrt{-\tilde{g}}\mathcal{L}_{m},
\end{equation}
this simplifies the action variation expression down to:
\begin{equation}
\delta S_m = \frac{1}{2} \int d^4x \sqrt{-g} \, T^{\mu\nu}_{m} \delta g_{\mu\nu} + \int d^4x \sqrt{-g} \, T_{m} \frac{d\ln \Omega}{d\sigma} \delta\sigma,
\end{equation}
where $T_{\mu\nu}\equiv \Omega^{6}\tilde{T}_{\mu\nu}$.
We apply an infinitesimal coordinate transformation $x^\mu \to x^\mu + \xi^\mu$. The field variations correspond to the Lie derivatives along the vector field $\xi^\mu$:
\begin{align}
\delta g_{\mu\nu} &= \nabla_\mu \xi_\nu + \nabla_\nu \xi_\mu, \\
\delta \sigma &= \xi^\mu \partial_\mu \sigma = \xi^\mu \nabla_\mu \sigma.
\end{align}
Substituting these geometric expressions back into the action variation reveals:
\begin{align}
\delta S_m &= \frac{1}{2} \int d^4x \sqrt{-g} \, T^{\mu\nu}_{m} \left( \nabla_\mu \xi_\nu + \nabla_\nu \xi_\mu \right) \notag\\ &+ \int d^4x \sqrt{-g} \, T_{m} \frac{d\ln \Omega}{d\sigma} \left( \xi^\nu \nabla_\nu \sigma \right).
\end{align}
Performing integration by parts on the first term to isolate $\xi^\nu$:
\begin{align}
\delta S_m = \int d^4x \sqrt{-g} \Big[ -\nabla_\mu T^{\mu}_{\phantom{\mu}\nu(m)} + \frac{d\ln \Omega}{d\sigma} T_{m} \nabla_\nu \sigma \Big] \xi^\nu
\end{align}
Thus the general coordinate covariance demands that $\delta S_M = 0$ for any arbitrary transformation vector field $\xi^\nu$ so that 
\begin{equation}
\nabla_\mu T^{\mu}_{\phantom{\mu}\nu(m)} = \frac{d\ln \Omega}{d\sigma} T_{m} \nabla_\nu \sigma.
\end{equation}
Matching this structural identity directly to the model's interaction term in \eqref{eq:apm} fixes the functional form of the conformal factor:
\begin{equation}
\frac{d\ln \Omega}{d\sigma} = \kappa \delta \implies \Omega(\sigma) = e^{\kappa \delta \sigma} \approx 1 + \kappa \delta \sigma
\end{equation}
To first linear order in the weak-field expansion, the physical metric tensor transforms as:
\begin{equation}
\tilde{g}_{\mu\nu} = (1 + 2\kappa \delta \sigma) g_{\mu\nu}.
\end{equation}

\bibliography{quintom}
\bibliographystyle{apsrev4-1} 

\end{document}